\documentclass[floats,floatfix,amssymb,prd,twocolumn,superscriptaddress,nofootinbib,preprintnumbers]{revtex4-1}

\usepackage{amssymb,amsmath,verbatim,mathtools,enumitem,etoolbox,graphicx,microtype,afterpage,bm}

\usepackage[utf8]{inputenc}
\usepackage{amsmath,amssymb}
\usepackage{amsfonts}
\usepackage{bm}
\usepackage{courier}
\usepackage{graphics}
\usepackage{float}
\usepackage{amsmath}
\usepackage{amssymb}
\usepackage[mathscr]{euscript}
\usepackage{float}
\usepackage[caption=false]{subfig}
\usepackage{lipsum}
\usepackage{mathtools}
\usepackage{multirow}
\usepackage{graphicx}
\usepackage{mathtools}
\usepackage{multirow}
\usepackage{graphicx}
\usepackage[usenames,dvipsnames,table,xcdraw]{xcolor}
\usepackage[colorlinks, pdfborder={0 0 0}]{hyperref}
\usepackage[nameinlink,capitalise]{cleveref}
\usepackage{tensor}
\usepackage{verbatim}
\usepackage{wrapfig}
\usepackage{soul} 

\graphicspath{{fig}}

\newcommand\aapr{A\&ARv}             
\newcommand\apjl{ApJ}                
\newcommand\araa{ARA\&A}             
\newcommand\mnras{MNRAS}             

\begin{document}
\title{Landscape of stellar-mass black-hole spectroscopy\\ with third-generation gravitational-wave detectors}

\author{Swetha Bhagwat}
\email{sbhagwat@star.sr.bham.ac.uk}
\affiliation{Institute for Gravitational Wave Astronomy $\&$ School of Physics and Astronomy, University of Birmingham,
Edgbaston, Birmingham B15 2TT, UK}

\author{Costantino Pacilio}
\email{costantino.pacilio@unimib.it}
\affiliation{Dipartimento di Fisica ``G. Occhialini'',
Universit\`a degli Studi di Milano-Bicocca, Piazza della Scienza 3, 20126 Milano, Italy}
\affiliation{INFN, Sezione di Milano-Bicocca, Piazza della Scienza 3, 20126 Milano, Italy}

\author{Paolo Pani}
\email{paolo.pani@uniroma1.it}
\affiliation{Dipartimento di Fisica, Sapienza Università
	di Roma, Piazzale Aldo Moro 5, 00185, Roma, Italy}
\affiliation{INFN, Sezione di Roma, Piazzale Aldo Moro 2, 00185, Roma, Italy}

\author{Michela Mapelli}
\email{michela.mapelli@unipd.it}
\affiliation{Physics and Astronomy Department Galileo Galilei, University of Padova, Vicolo dell’Osservatorio 3, I–35122, Padova, Italy}
\affiliation{INFN - Padova, Via Marzolo 8, I–35131 Padova, Italy}
\affiliation{INAF - Osservatorio Astronomico di Padova, Vicolo dell’Osservatorio 5, I-35122 Padova, Italy}

\begin{abstract}
Gravitational-wave black-hole spectroscopy provides a unique opportunity to test the strong-field regime of gravity and the nature of the final object formed in the aftermath of a merger.
Here we investigate the prospects for black-hole spectroscopy with third-generation gravitational-wave detectors, in particular the Einstein Telescope in different configurations, possibly in combination with Cosmic Explorer.
Using a state-of-the-art population model for stellar-origin binary black holes informed by LIGO-Virgo-KAGRA data, we compute the average number of expected events for precision black-hole spectroscopy using a Fisher-matrix analysis.
We perform our analysis on the dominant mode $(2,2,0)$ and a set of subdominant modes $[(3,3,0), (2,1,0), (4,4,0)]$ using amplitude and phase fits corresponding to the aligned spin configurations.
We find that Einstein Telescope will measure two independent quasinormal modes within ${\cal O}(1)\%$ (resp. ${\cal O}(10)\%$) relative uncertainty for at least ${\cal O}(1)$ (resp. ${\cal O}(500)$) events per year, with similar performances in the case of a single triangular configuration or two L-shaped detectors with same arm length. A 15-km arm-length configuration would improve rates by roughly a factor of two relative to a 10-km arm-length configuration. When operating in synergy with Cosmic Explorer the rates will improve significantly, reaching few-percent accuracy for ${\cal O}(100)$ events per year.
\end{abstract}

\preprint{ET-0106A-23}

\maketitle

\section{Introduction}

When a distorted black hole~(BH) forms in a binary BH merger, it relaxes emitting gravitational waves~(GWs). Sufficiently later after the merger, the evolution of the remnant spacetime can be described by perturbation theory and the GW signal emitted during this phase is called the ringdown~\cite{QNM-Chandrasekhar,Vishveshwara}.

At asymptotic infinity, the ringdown signal can be analytically approximated as a linear superposition of countably infinite damped sinusoids corresponding to a characteristic frequency and damping time spectra called the quasinormal modes~(QNMs) (see~\cite{Kokkotas-review,Berti:2009kk,Konoplya:2011qq} for some reviews). BH spectroscopy~\cite{Berti:2005ys,gossan-et-al,RD-TGR-first-proposal} --namely the measurement of the frequencies and damping times in a ringdown signal~-- allows us to perform multiple null tests of the General theory of Relativity~(GR) and of the nature of the BHs~\cite{nollert1999quasinormal,Berti:2018vdi,PhysRevD.85.082003,Maselli:2019mjd,Volkel:2020daa,Carullo:2018sfu,Cardoso:2019rvt}.

GR predicts that the QNM spectrum of the BH remnant can be fully parameterized by its mass and (dimensionless) spin. The measurement of the parameters of a single complex QNM can be inverted to obtain an estimate of the mass and spin of the BH, whereas with the measurement of additional QNMs one can perform consistency tests for the Kerr nature of the source, including tests of the no-hair and area theorems~\cite{No-hair-original,Isi:2020tac}.

Even in those cases in which the remnant is a Kerr BH, and therefore the QNM spectrum is consistent with the GR prediction, the ringdown amplitudes and phases can be used to test GR and the nature of the progenitor binary~\cite{Forteza:2022tgq}.

Since the onset of GW observations in 2015, significant effort has been devoted to measuring the ringdown in the LIGO-Virgo-KAGRA merger events. The fundamental QNM has been measured for GW150914 \cite{Carullo:2019flw,TGR-gw150914,Brito:2018rfr}. Moreover, inferences of the final mass and spin from the ringdown are routinely performed \cite{LIGOScientific:2020tif,LIGOScientific:2021sio} and they are checked for consistency with estimates from the inspiral-merger part of the signals.
There has been no unambiguous and confident measurements of any secondary QNM parameters in the events detected thus far, although exciting hints of overtones have been found in GW150914~\cite{Isi:2019aib,Finch:2022ynt} and the measurement of a secondary angular QNM has been reported in GW190521~\cite{Capano:2021etf,Capano:2022zqm}.
However, BH spectroscopy with overtones may be particularly affected by potential limitations related to the resolvability of the QNMs~\cite{Bhagwat:2019dtm,Ota:2019bzl}, the number of overtones that need to be included for an unbiased parameter recovery~\cite{Isi:2019aib,Giesler:2019uxc}, the physical interpretation of the measurement including the risk of over-fitting~\cite{Cotesta:2022pci,Baibhav:2023clw}, and sensitivity to the choice of the start time of the ringdown~\cite{Cotesta:2022pci,2020PhRvD.101d4033B,JimenezForteza:2020cve,Ota:2019bzl,Nee:2023osy} owing to the short decay times. Furthermore, merger events detected by current interferometers are expected to have a low signal-to-noise ratio~(SNR) in the secondary angular mode~\cite{Capano:2021etf,Berti:2016lat}, limiting the constraining power of such tests.

Next-generation GW detectors have the potential to perform \emph{precision BH spectroscopy}, allowing for sub-percentage accuracy tests. In an earlier paper~\cite{Bhagwat:2021kwv}, we discussed the landscape of BH spectroscopy using supermassive BH binary mergers detectable with the future space mission LISA~\cite{2017arXiv170200786A} and with third-generation ground-based detectors such as the Einstein Telescope~(ET)~\cite{2010CQGra..27s4002P,Maggiore:2019uih,Reitze:2019iox, Kalogera:2021bya, Hild:2010id}. In this study, we present a similar analysis of the landscape of BH spectroscopy for the stellar-mass binary BH~(BBH) population -- the prime candidates for ET~\cite{Hild:2010id,Maggiore:2019uih,Kalogera:2021bya}. We study the prospects of BH spectroscopy for both individual events and for population studies using a state-of-the-art population model that is informed by the LIGO-Virgo-KAGRA observations~\cite{mapelli2016,perigois2023}. We consider two of the most relevant configurations adopted in the recent analysis on the science impact of different ET designs~\cite{CoBA}. Lastly, we also study the prospects for joint ringdown measurement with ET and the Cosmic Explorer~(CE)~\cite{Evans:2016mbw,Essick:2017wyl} which is an American initiative for a third-generation ground-based GW detector.

The remainder of this paper is organized as follows. In Sec.~\ref{sec:setup}, we outline our setup for BH spectroscopy. Then, in Sec.~\ref{sec:golden}, we study BH spectroscopy for golden events, i.e., loud events similar to GW150914. In Sec.~\ref{sec:pop}, we discuss the properties of the stellar-mass BBH population in the context of ringdown tests, and present the results of ringdown population studies with ET. Next in Sec.~\ref{sec:ETCE}, we study the advantages of a joint detection with ET and CE. Finally, we conclude our work with a discussion and future research directions in Sec.~\ref{sec:conclusions}.

\section{Analysis setup\label{sec:setup}}
Within linear perturbation theory \cite{Teukolsky:1972my,teukolsky1973perturbations}\footnote{See Refs.~\cite{Cheung:2022rbm,Mitman:2022qdl,Lagos:2022otp,Kehagias:2023ctr,Kehagias:2023mcl,Baibhav:2023clw} for recent studies about next-to-leading order ringdown effects. Note that the latter are particularly relevant for the 440 mode (which can be sourced by the fundamental 220 mode at the quadratic order), whereas the 220, 330, and 210 modes are less contaminated by nonlinear effects.} , the ringdown signals can be analytically modeled as $h(t)=h_+(t)+ih_\times(t)$, where
\begin{equation}
\label{eq:rdmodel}
\begin{split}
    &h_{+}(t)=\sum_{lmn}{\cal A}_{lmn}\cos\left(2\pi f_{lmn}t+\phi_{lmn}\right)e^{-t/\tau_{lmn}}\,\mathcal{Y}^{lm}_{+}(\iota)\,,
    \\
    &h_\times(t)=\sum_{lmn}{\cal A}_{lmn}\sin\left(2\pi f_{lmn}t+\phi_{lmn}\right)e^{-t/\tau_{lmn}}\,\mathcal{Y}^{lm}_{\times}(\iota).
\end{split}
\end{equation}
where $\iota$ is the remnant spin inclination angle and
$ \{f_{lmn}, \tau_{lmn}, \mathcal{A}_{lmn}, \phi_{lmn} \}$ are the frequency, damping time, amplitude, and phase of the $(lmn)$ QNM, respectively. The integers $(lmn)$ refer to the multipolar, azimuthal, and overtone index, respectively, where $n=0$ corresponds to the fundamental tone and $(lmn)=(220)$ is the dominant mode in a quasicircular coalescence.
The excitation amplitudes roughly scale as ${\cal A}_{lmn}\sim M_f/d_L$, where $M_f$ is the mass of the remnant BH and $d_L$ is the luminosity distance from the source \cite{Gossan:2011ha}.
The $\{+,\times\}$ polarizations should be defined using a spin-weighted spheroidal harmonics basis. However, for simplicity we approximate it with their spherical harmonic counterpart, which is a good approximation for moderately spinning remnants (see Refs.~\cite{London:2018nxs,Cook:2020otn,Baibhav:2023clw} for related discussion). The polarization basis can be written as~\cite{q-from-rd,Berti:2007zu}
\begin{equation}
    \mathcal{Y}_{+, \times}(\iota) = \tensor[_{-2}]{\mathcal{Y}}{^{lm}}(\iota,0) \pm (-1)^l\tensor[_{-2}]{\mathcal{Y}}{^{l-m}}(\iota,0)\,.
\end{equation}
Note that Eq.~(1) assumes equatorial reflection symmetry in the mode excitations, for which $\mathcal A_{lm}e^{i\phi_{lm}} = (–1)^l \mathcal A_{l–m}e^{-i\phi_{l-m}}$. As a consequence, the ellipticity (polarization ratio) of all observed modes will be a function of a single parameter (the inclination angle), as prescribed by the spherical harmonic basis \cite{isi2021analyzing}.

In this work, we consider the performance of BH spectroscopy with angular QNMs, since they have longer damping times and they can be reliably extracted sufficiently after the peak, where linear perturbation theory is accurate, unlike the case of overtones~\cite{Bhagwat:2019dtm,Ota:2019bzl,JimenezForteza:2020cve,Baibhav:2023clw,Nee:2023osy}.

Using the above template, we infer the statistical errors associated to the measurements of the ringdown parameters in a Fisher information matrix framework. In particular, we use the same analysis setup developed in~\cite{Bhagwat:2021kwv} (to which we refer the readers for further technical details). The setup is based on a fully numerical evaluation of the Fisher matrix.

Using the above template, we infer the statistical errors associated to the measurements of the ringdown parameters in a Fisher information matrix framework. The setup is based on a fully numerical evaluation of the Fisher matrix similar to that in \cite{Bhagwat:2021kwv}; our formalism differs from \cite{Berti:2005ys} in that we consider all modes simultaneously without averaging over inclination angle and over pattern functions as well as we do not assume the large $\tau_{lmn}$ limit, i.e., we use the full Lorentzian in the frequency domain instead of $\delta$-functions peaked at $f_{lmn}$. We verified that all the Fisher matrices used in our computations are invertible without significant numerical error in the inversion.

We find it convenient to parameterize the frequencies and damping times as
\begin{equation}
f_{lmn}=f_{lmn}^{\rm Kerr}(1+ \delta f_{lmn})\,, \quad
\tau_{lmn}=\tau_{lmn}^{\rm Kerr}(1+ \delta \tau_{lmn})\,,
\end{equation}
where $f_{lmn}^{\rm Kerr}$ and $\tau_{lmn}^{\rm Kerr}$ are the GR-predicted frequencies and damping times of a remnant Kerr BH; these are functions of the final mass $M_{f}$ and the final spin $\chi_{f}$. The (dimensionless) deviation parameters $\delta f_{lmn}$ and $\delta \tau_{lmn}$ quantify the departure of the measured spectrum from the GR prediction.
For GR to pass a null-hypothesis test with a certain level of confidence, the inferred posterior distributions of $\delta f_{lmn}$ and $\delta \tau_{lmn}$ must contain zero within a given confidence level. 
Note that, in the absence of extra information coming from the inspiral, $\delta f_{220}$ and $\delta \tau_{220}$ are degenerate with $M_f$ and $\chi_f$ \cite{isi2021analyzing,Pacilio:2023mvk}. Therefore, without loss of generality we use ($M_f,\chi_f$) in place of ($\delta f_{220},\delta \tau_{220}$), and we use $\delta f_{lmn}$ and $\delta \tau_{lmn}$ for any other mode.
Overall, the parameters of our Fisher matrix are 
\begin{equation}
\{M_{f},\chi_{f},\mathcal{A}_{220},\phi_{220},\delta f_{lmn}, \delta \tau_{lmn}, \mathcal{A}_{lmn}, \phi_{lmn}\}\,,
\end{equation}
where $(lmn)\neq(220)$.
Note that the template has $16$ parameters.

The injected amplitudes $\mathcal{A}_{lmn}$ are obtained from the properties of the progenitor binary (and therefore assuming GR). The injected amplitude $\mathcal{A}_{220}$ of the fundamental mode is taken from~\cite{Gossan:2011ha}, while the amplitudes of the other modes relative to $\mathcal{A}_{220}$ are obtained using the recent fits in Ref.~\cite{Forteza:2022tgq}. 
We also inject relative amplitudes $\phi_{lmn}-\phi_{220}$ consistently with the fits in \cite{Forteza:2022tgq}, while $\phi_{220}$ is randomly sampled within $[0,\pi]$. The amplitude and phase fits start $\sim 10 M$ after the peak of strain and the SNR integration limits are consistent with this choice.

The injected values of the final spin $\chi_f$ are derived from the mass ratio $q$ using the numerical fits in~\cite{Hofmann:2016yih}. We stress that the fits of Refs.~\cite{Forteza:2022tgq, Hofmann:2016yih} are only needed to select realistic injected values, but $M_f$, $\chi_f$, and the QNM amplitudes and phases are free parameters of the waveform. We assume a GR signal and inject $\delta f_{lmn}=0=\delta \tau_{lmn}$.

\begin{table*}[t]
    \centering
    \begin{tabular}{||c|c|c|c|c|c|c|c|c|c|c|c|c||}
         \hline\hline
         $q$ & $\rho_{\rm RD}$ & $\sigma(M_f)/M_f$ & $\sigma(\chi_f)$ & $\sigma(\delta f_{330})$ & $\sigma(\delta f_{210})$
         & $\sigma(\delta f_{440})$ & $\sigma(\delta \tau_{330})$ & $\sigma(\delta \tau_{210})$ & $\sigma(\delta \tau_{440})$ & $\sigma(\mathcal{A}_{330})/\mathcal{A}_{330}$ & $\sigma(\mathcal{A}_{210})/\mathcal{A}_{210}$ & $\sigma(\mathcal{A}_{440})/\mathcal{A}_{440}$\\
         \hline\hline
         1.2 & 84 & 0.03 & 0.04 & 0.03 & 0.18 & 0.03 & 0.60 & 1.08 & 0.63 & 0.48 & 1.86 & 0.46\\
         2 & 85 & 0.04 & 0.06 & 0.01 & 0.05 & 0.03 & 0.23 & 0.32 & 0.64 & 0.17 & 0.53 & 0.48\\
         5 & 64 & 0.08 & 0.17 & 0.02 & 0.06 & 0.04 & 0.29 & 0.32 & 0.51 & 0.23 & 0.41 & 0.48\\
         \hline\hline
    \end{tabular}
\caption{Estimated $1$-$\sigma$ errors on the QNM parameters for a GW150914-like system with $M_{f}  = 70 M_{\odot}$ at luminosity distance $d_L=450\,{\rm Mpc}$ with $({\rm ra,dec},\psi,\iota)=(1.16,-1.19,1.12,\pi/3)$ and occurring at a geocenter GPS time $t_{\rm GPS}=1126259466.43$. The reported ringdown SNR, $\rho_{\rm RD}$, refers to a triangular ET detector placed in Sardinia with ET-D sensitivity. Within the assumptions of the Fisher matrix, absolute errors scale inversely with the SNR.
}
\label{tab:typical-system}
\end{table*}

\section{BH Spectroscopy with ET for Golden events} \label{sec:golden}
In this section, we investigate the performance in BH spectroscopy on isolated ``golden" (i.e., loud and favorable) events using ET. We consider a GW150914-like system with $M_{f} = 70 M_{\odot}$ at a luminosity distance $d_L=450\, {\rm Mpc}$.
We place the event at the sky position ${\rm(ra,dec)}=(1.16,-1.19)$ with polarization $\psi=1.12$, and assume it occurs at a geocenter GPS time $t_{\rm GPS}=1126259466.43$. In order to ensure that the subdominant modes are not suppressed, we choose an inclination angle $\iota=\pi/3$.

We inject the mode amplitudes and phases as explained above.
Since the mass ratio $q$ has a leading effect on the amplitude excitation factors compared to the binary spins~\cite{London:2018nxs,Kamaretsos:2012bs,gossan-et-al,Forteza:2022tgq}, here we consider three different values of $q \in \{1.2, 2, 5 \}$ for nonspinning systems. The corresponding remnant spins as obtained from \cite{Hofmann:2016yih} are $\chi_f\in\{0.68,0.62,0.42\}$.

We tabulate the results of our Fisher matrix analysis in Table~\ref{tab:typical-system}.
Since the results have a simple scaling with the SNR within the approximation of the Fisher-matrix analysis, in this section we consider ET in a triangular configuration adopting the standard ET-D sensitivity curve~\cite{Hild:2010id}.
We see that ET ringdown signals will have about one order of magnitude higher SNR compared to the same events detected by LIGO-Virgo. This roughly translates into an order-of-magnitude improvement on the measurement errors in the ringdown parameters. Note that for the same final mass and luminosity distance, the ringdown SNR ${\rho_{\rm RD}}$ decreases significantly with $q$; this is unfortunate for BH spectroscopy because the excitation of the subdominant angular modes is suppressed for almost equal-mass binaries.
At the same time, \textit{for a fixed SNR}, the uncertainty of the QNM parameters tends to decrease with increasing $q$ as the amplitude ratios of the subdominant modes increase. Thus, as $q$ increases there is an interplay between decreasing ringdown SNR and increasing $\mathcal{A}_{lmn}$ that produces the trend in Table~\ref{tab:typical-system}.

Overall, we see that ET can measure the subdominant QNM frequencies within a few percent accuracy, with $\delta f_{330}$ always being the best constrained subdominant parameter. For systems with $q \geq 2$, ET will measure $\tau_{330}$ and $\tau_{210}$ within $20\%$ to $30\%$ accuracy, while measuring $\tau_{440}$ is more difficult, with about $50\%$ to $65\%$ uncertainties in its recovery. Finally, although the recovery of amplitude ratios is not directly relevant for performing BH spectroscopy, it can be used to quantify the confidence level in the detection of a secondary mode (if $\sigma(\mathcal{A}_{lmn})/\mathcal{A}_{lmn}\ll 1$) and to perform other tests of GR, such as the recently-proposed amplitude-phase consistency test~\cite{Forteza:2022tgq} and merger-ringdown test~\cite{Bhagwat:2021kfa}. Thus, we report that, for $q \geq 2$, ET can measure $\mathcal{A}_{330}$ with $\approx 20\%$ accuracy, whereas it can measure $\mathcal{A}_{210}$ and $\mathcal{A}_{440}$ with $\approx 50\%$ accuracy.
\section{BH Spectroscopy with ET with a stellar mass BBH population}
\label{sec:pop}
%
Having discussed the prospect for BH spectroscopy with single golden events, in this section we move to the prospects for BH spectroscopy using a realistic stellar-mass BBH population.

\subsection{Stellar-mass BBH population model}
\label{sec:pop:1}

We use the BBH population described in~\cite{mapelli2022}. In particular, we generated a population of merger signals by assuming that the BBHs observed by the LIGO--Virgo--KAGRA Collaboration come from a mixture of the dynamical and isolated channel. The isolated channel consists of BBHs that evolve from unperturbed massive binary stars~\cite{bethe1998,belczynski2002}. We evolved our massive binary systems with the population-synthesis code {\sc mobse}~\cite{mapelli2017,giacobbo2018}, which implements an up-to-date model for stellar winds~\citep{giacobbo2018b}, and a formalism for the main binary evolution processes~\citep{hurley2002}. We describe the outcome of core-collapse supernovae with the rapid model by~\cite{fryer2012}, which enforces a  mass gap between $2\,M_\odot$ and $5\,M_\odot$~\cite{oezel2010,farr2011}. We also account for (pulsational) pair instability, as described in~\cite{mapelli2020}. In our models, pair instability produces a mass gap between $\approx{60\,M_\odot}$ and $\approx{120\,M_\odot}$ M$_\odot$~\citep{giacobbo2018}. This gap is partially filled by dynamically formed binaries described below. We generate the dimensionless spin magnitudes $\chi{}$ of the isolated BBHs from a Maxwellian distribution with root-mean-square parameter $\sigma_{\chi}=0.1$ and truncated at $\chi=1$. This is a toy model that allows us to reproduce the main features of BBHs observed by LIGO--Virgo--KAGRA after the third observing run~\citep{popandrateO3}. We choose a toy model for spin magnitudes, because of the large uncertainties still affecting astrophysical models and hampering their predictive power~\citep{perigois2023}. We assume that binary evolution processes align the spins of the progenitor stars with the orbital angular momentum of the binary system: only the supernova explosion can produce a misalignment between the BH spin and the orbital angular momentum~\citep{rodriguez2016b}. This results in a preference for aligned spins in our isolated BBHs. 

According to the dynamical channel, BBHs assemble in dense stellar clusters by three-body encounters and dynamical exchanges (e.g., \cite{portegieszwart2000,banerjee2010,mapelli2016,rodriguez2016,fragione2018,kremer2020,banerjee2021}). We model three different astrophysical populations of star clusters: nuclear, globular, and young  clusters. Nuclear clusters lie at the center of their host galaxies and can be very massive ($\approx{10^7}\,M_\odot$,~\cite{neumayer2020}). Globular clusters are massive ($\approx{10^{4-6}\,M_\odot}$,~\cite{harris2010}) and mostly formed in the early Universe ($z\sim{2-4}$,~\cite{vandenberg2013}), while young clusters are less massive than the other two families ($\leq{10^5}$M$_\odot$) and are one of the most common birthplaces of massive stars in the local Universe~\citep{portegieszwart2010}. We model the star formation history of nuclear, globular, and young clusters as described in~\cite{mapelli2022}. In our model, dynamically assembled BBHs in nuclear, globular, and young clusters can undergo hierarchical mergers~\citep{miller2002,gerosa2017,fishbach2017,rodriguez2019,antonini2019,antonini2022,gerosa2021}: if the remnant of the merger of two stellar-origin BHs is retained inside its parent cluster despite the gravitational recoil~\citep{merritt2004,campanelli2007}, it can pair up again with another BH and lead to a second-generation (or nth-generation) merger.

We generate the masses of first-generation BHs from the {\sc mobse} population synthesis code, i.e., the same code we use for the isolated binaries, for consistency. We also randomly draw the spins of first-generation BHs from the same Maxwellian distribution as we described for isolated BBHs. The masses  and the spins of second-generation BHs are obtained with fitting formulas to numerical relativity simulations~\citep{jimenezforteza2017}. This means that the spin magnitudes of our second-generation BHs peak at $\chi{}\sim{0.7-0.8}$. Finally, the spins of both first-generation and second-generation dynamical BHs are isotropically oriented over the sphere, accounting for the effect of dynamical encounters~\cite{rodriguez2016b}.

With this set up, we obtain a primary BH mass distribution (Fig.~6 of~\cite{mapelli2022})  and local BBH merger rate density ($R\approx{31}$ Gpc$^{-3}$ yr$^{-1}$) that lie within the 90\% credible intervals inferred by the LIGO--Virgo--KAGRA Collaboration after the third observing run. We use this catalog because it is grounded on state-of-the-art astrophysical models and matches the main observed features.   We refer to~\cite{mapelli2022} for more details on our simulations. Out of these simulations, we randomly extract a   sub-sample comprising all BBHs that we expect to merge in a time span of 10 years, from the local Universe out to redshift $z=14$. This yields  a final catalog of 1181195 BBHs.

\subsection{Population distributions in the context of BH spectroscopy}
Before discussing our results it is useful to investigate some general properties of our stellar-mass population model in the context of BH spectroscopy.

\begin{figure*}[th]
    \centering
    \includegraphics[width=0.375\textwidth]{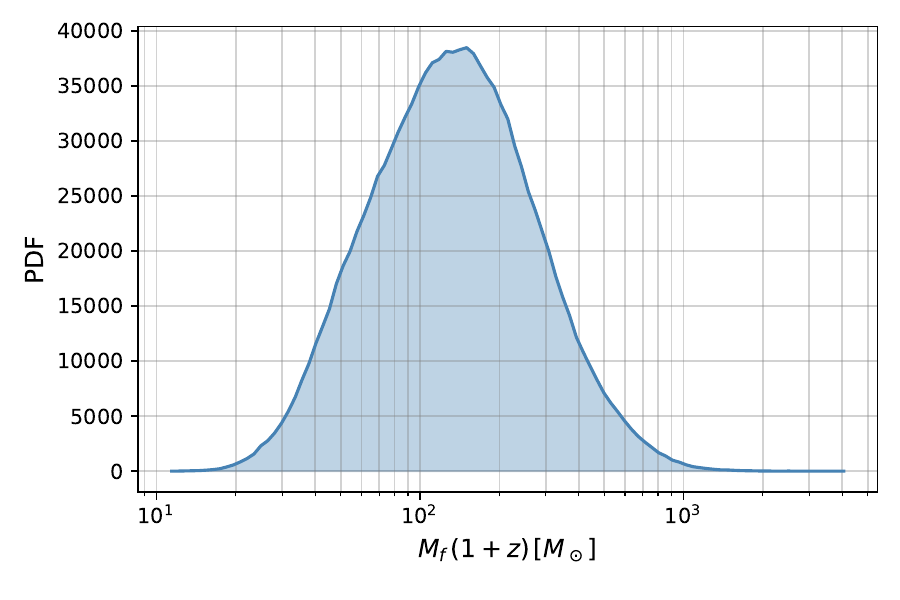}
    \includegraphics[width=0.375\textwidth]{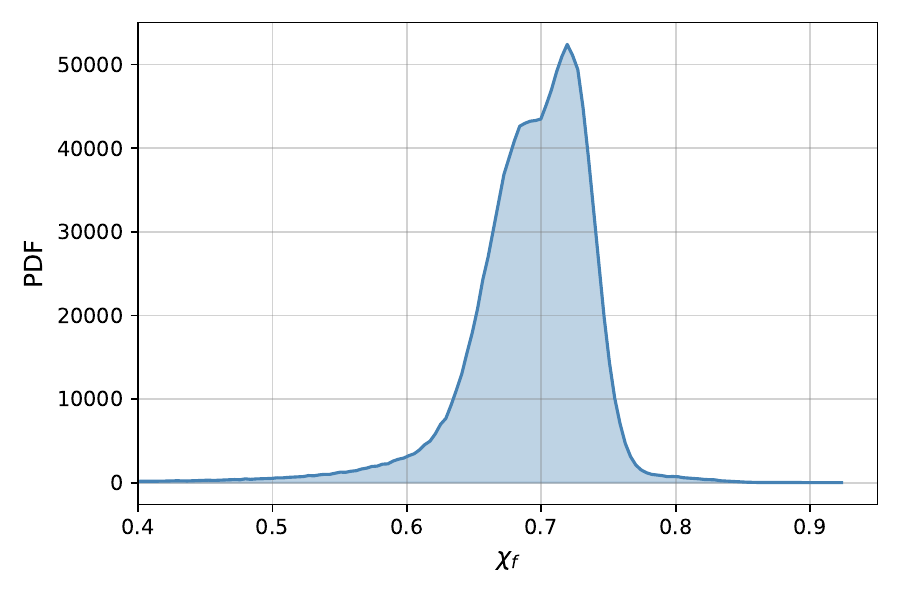}\\
    \includegraphics[width=0.375\textwidth]{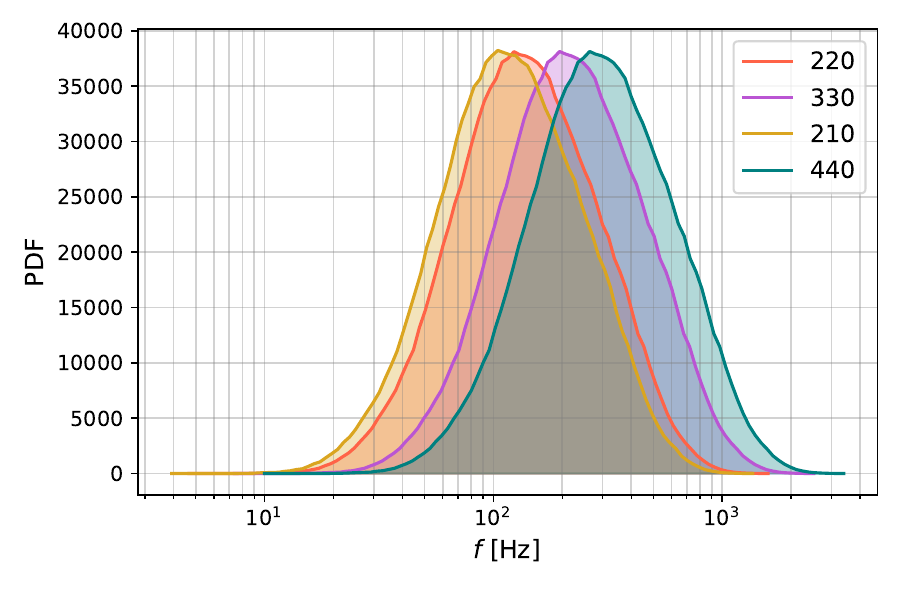}
    \includegraphics[width=0.375\textwidth]{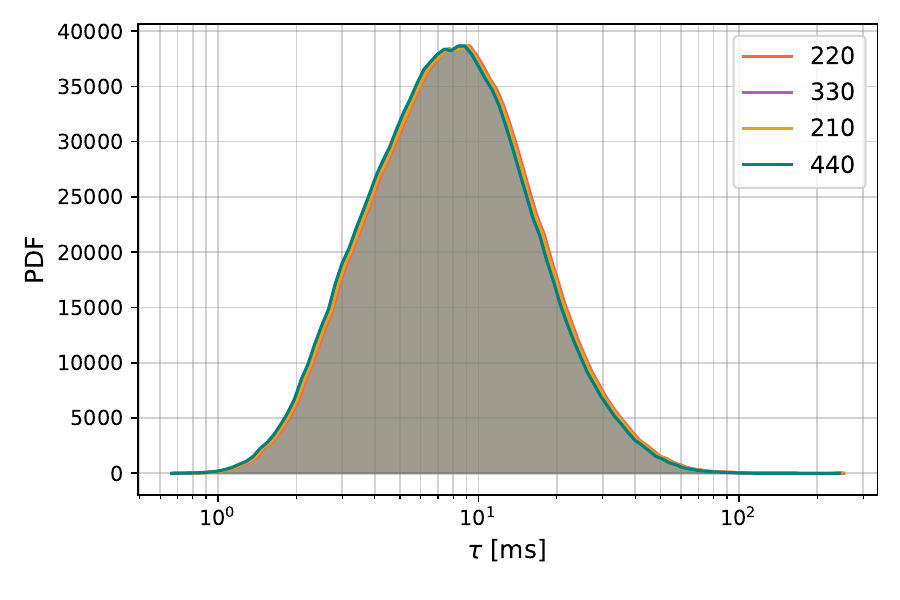}\\
    \includegraphics[width=0.375\textwidth]{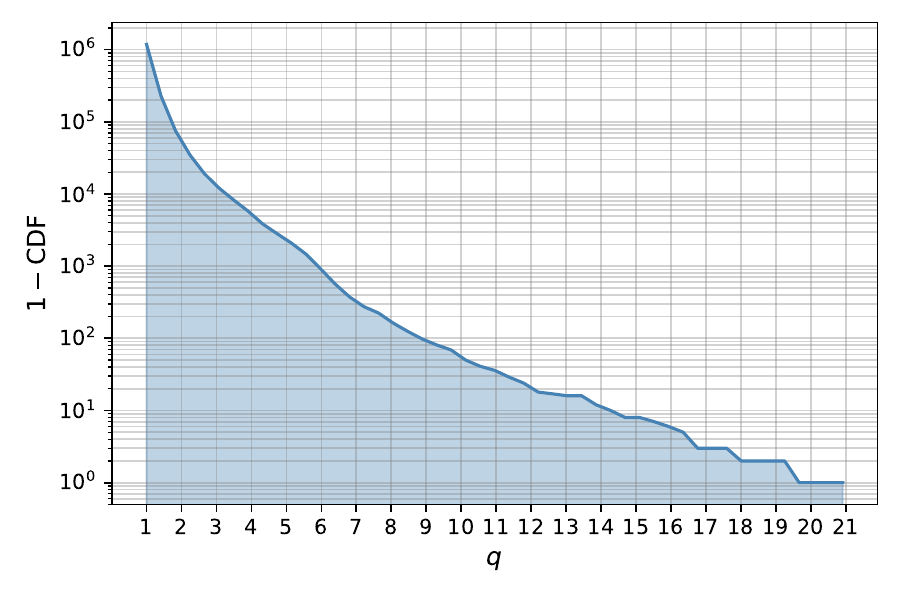}
    \includegraphics[width=0.375\textwidth]{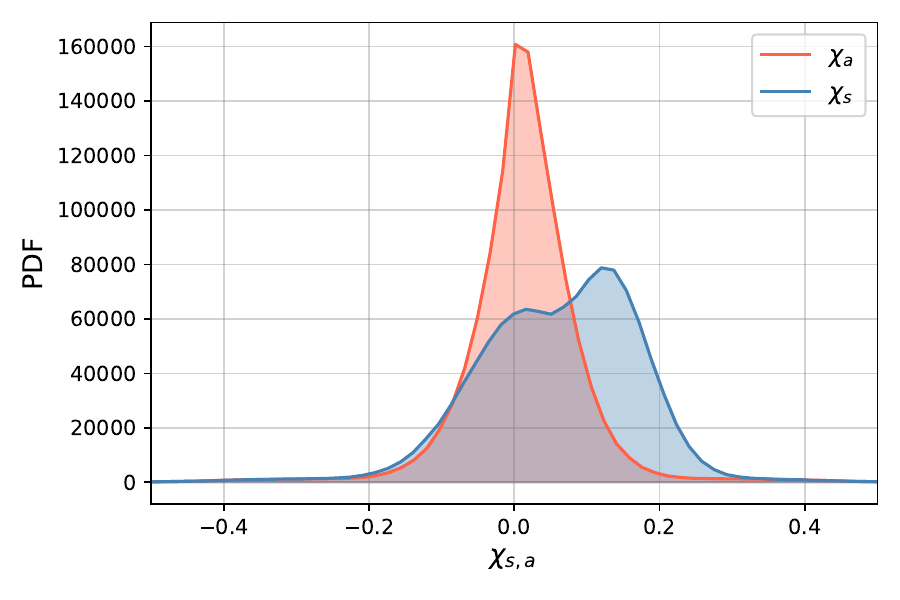}\\
    \caption{Distributions of
     remnant detector-frame mass $M_f\,(1+z)$ (top left) and dimensionless spin $\chi_f$ (top right), the corresponding
     QNM frequencies $f_{lmn}^{\rm Kerr}$ (middle left) and damping times $\tau_{lmn}^{\rm Kerr}$ (middle right), and mass ratio $q$ (bottom left), and symmetric spin $\chi_s$ and anti-symmetric spin $\chi_a$ components (bottom right) of the progenitors.
    All  histograms are not normalized and show the actual number of events for the full 10-yr catalog.
    }
\label{fig:Freq-tau-22}
\end{figure*}

The top panel in Fig.~\ref{fig:Freq-tau-22} shows the distributions of the detector-frame masses and of the final spin of the BH remnant. These distributions can be mapped into the characteristic frequencies and damping times of the QNMs assuming a Kerr BH, as presented in the middle panels of Fig.~\ref{fig:Freq-tau-22} for various subdominant modes.
If detector detuning is viable for next-generation ground-based detectors, then the middle-left panel can be used to infer the frequency ranges that optimize the detection of various subdominant modes.  In the middle-right panel, we see that all modes have a very similar damping time distribution, peaking at approximately $10\,{\rm ms}$. These ranges can provide informed priors in a Bayesian parameter estimation.

A notable property of the catalog in the context of BH spectroscopy is the distribution of the binary mass ratio and of the spins. The asymmetry in the progenitor binary systems is given by the mass ratio $q$ and the initial BH spins. Note that the greater the asymmetry in the progenitor BBH, the higher the subdominant mode excitation~\cite{London:2014cma,Kamaretsos:2012bs,gossan-et-al,JimenezForteza:2020cve,Forteza:2022tgq}. 
In the bottom panels of Fig.~\ref{fig:Freq-tau-22} we show the inverse cumulative distribution, $1-{\rm CDF}$, of $q$ and the probability distribution of the spin combinations $\chi_{s,a}=(m_1\chi_{1,z}\pm m_2\chi_{2,z})/(m_1+m_2)$ for a 10-year catalog. Although the distribution of $q$ peaks at equal mass binaries, we see that there is a significant number of events with $q\neq1$. For these systems the subdominant QNM excitation is nonnegligible. For instance, we expect $\sim 10^3$ event/yr with $q\geq3$ and $\sim 10^2$ events/yr with $q\geq6$.
Furthermore, even for $q\approx 1$, the subdominant QNM excitations can be triggered by the spins of the progenitor BBHs.
We see that $\chi_{a}$ peaks close to zero with support in the range $ \pm 0.2$, whereas $\chi_{s}$ favors a positive value with a peak around 0.18.

\begin{figure}[th]
\includegraphics[width=0.4\textwidth]{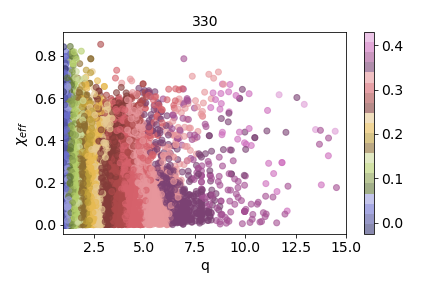} \\
\includegraphics[width=0.4\textwidth]{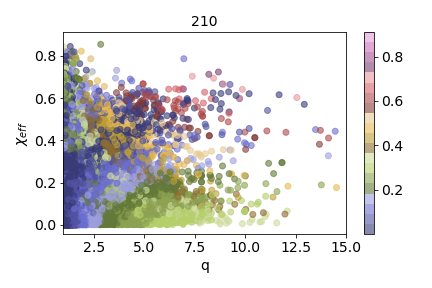} \\
\includegraphics[width=0.4\textwidth]{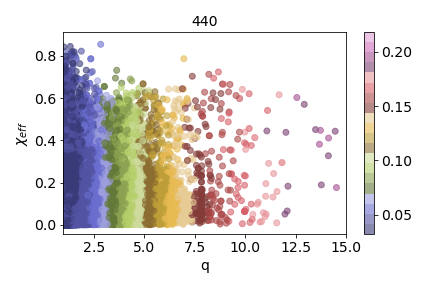}
\caption{Scatter plot of the amplitude ratio $\mathcal{A}_{lmn}/\mathcal{A}_{220}$ on the $q$-$\chi_{\rm eff}$ plane for the entire $10\,{
\rm yr}$ catalog. Here $\chi_{\rm eff}=(m_{1} \chi_{1}+m_{2} \chi_{2})/(m_{1}+m_{2})$ with $m_{i}$ denoting the progenitor masses and $\chi_{i}$ progenitor spin. In each subplot, the color bar displays $\mathcal{A}_{lmn}/\mathcal{A}_{220}$ of the $330$ mode (top), the $210$ mode (center) and the $440$ mode (bottom).}
\label{fig:assym}
\end{figure}

Using the fits\footnote{Note that the fits in \cite{Forteza:2022tgq} assume aligned spins and, therefore, we used the spin components along the z-direction, $\chi_{1,z}$ and $\chi_{2,z}$, to compute the injected values of the amplitude ratios ${\cal A}_{lmn}/{\cal A}_{220}$.} of~\cite{Forteza:2022tgq}, we can map these distributions into that of the amplitude ratio $\mathcal{A}_{lmn}/\mathcal{A}_{220}$, as in the scatter plot in Fig.~\ref{fig:assym}, where the color bar shows the magnitude of the mode excitation. For 330 and 440 the spin contribution is much less important than that for 210~\cite{Forteza:2022tgq}.
Indeed, the amplitude ratio for 210 spans a larger range owing to its stronger dependence on the progenitor spins. Note that these distributions of the amplitude ratios of the various subdominant modes can be used to choose informed priors when performing Bayesian parameter estimation.

\begin{figure*}[th]
    \centering
    \includegraphics[width=0.415\textwidth]{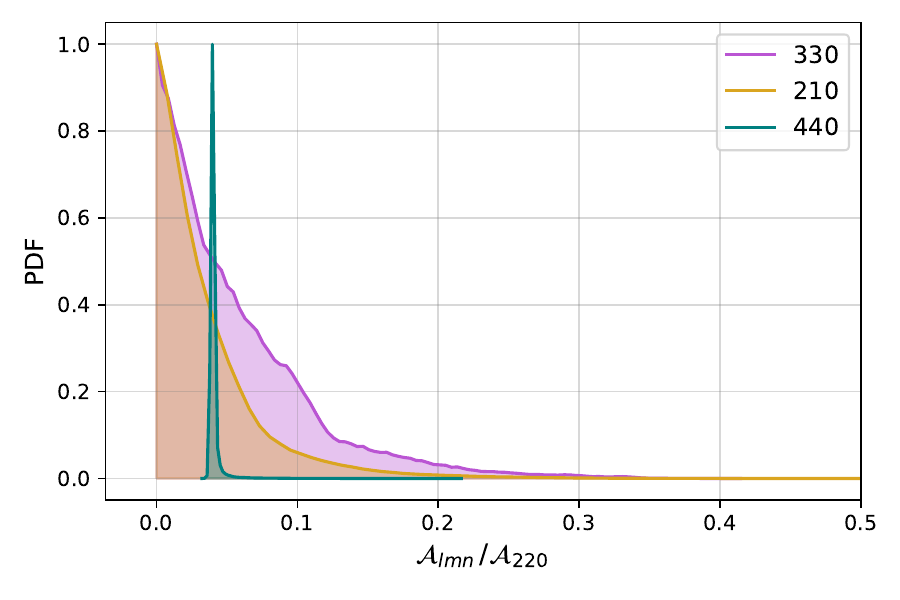}
    \includegraphics[width=0.415\textwidth]{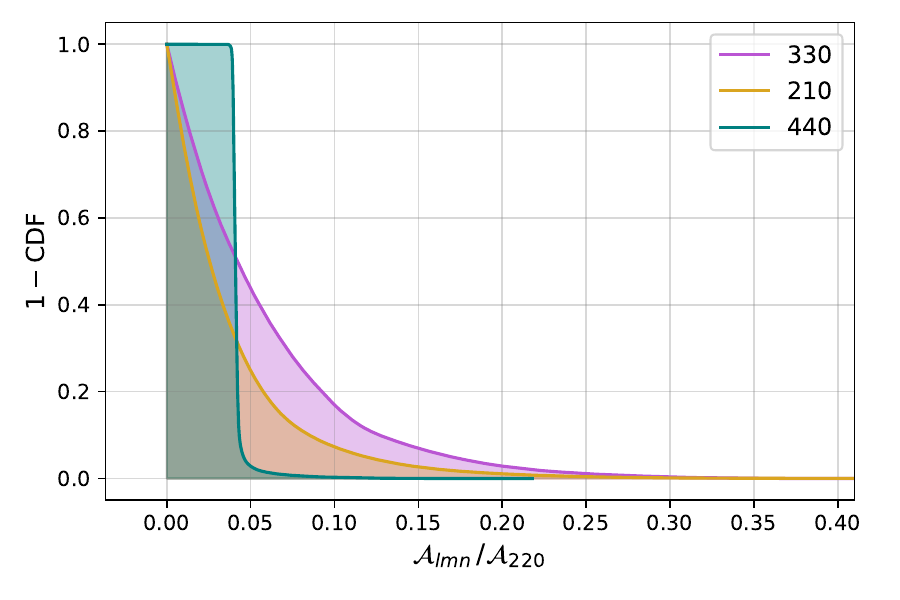}
    \caption{Unnormalized probability distributions (left) and inverse cumulative distributions (right) of the amplitude ratios $\mathcal{A}_{lmn}/\mathcal{A}_{220}$ for the excitation modes considered in this work.
    }
    \label{fig:amps}
\end{figure*}

Finally, in Fig.~\ref{fig:amps}, we present the (unnormalized) probability distribution (left) and the inverse cumulative distribution (right) of the amplitude ratios. We see that $\sim 20 \%$ and $\sim 10 \%$ of the events have $\mathcal{A}_{330} \sim 0.1$ and $\mathcal{A}_{221} \sim 0.1$, respectively. While the amplitude distributions of $330$ and $210$ are comparable, the peak of the amplitude ratio of $440$ is very sharp at $\mathcal{A}_{440}\sim 0.04$ so that it is very unlikely to find events with larger $\mathcal{A}_{440}$.

\subsection{Landscape of BH spectroscopy with ET} \label{subsec:landscapeET}

In this section we present the landscape of BH spectroscopy for ringdowns of stellar mass BH mergers with ET. We consider the most relevant configurations adopted in the recent~\cite{CoBA}. We focus on the `hybrid' cryogenic configuration comprising of a high-frequency~(HF) and a low-frequency~(LF) instruments (which updates the standard ET-D curve and was labelled `HFLF' in~\cite{CoBA}). We did not find any significant difference when the low-frequency (LF) instrument is absent. This is due to the fact that the ringdown modes can be represented as a Lorentzian, which is narrow for slow damping and, in our catalog, there exists a negligible fraction of events with $f_{220}<20\,{\rm Hz}$ (see Fig.~\ref{fig:Freq-tau-22}), where the contribution of the LF instrument improves the sensitivity curve.
While we found that the detector geometry (either triangular or L-shaped) does not significantly impact the performances of BH spectroscopy, the detector arm-length is more relevant.
Therefore, we present the result for two representative configurations: 1) a single, 10-km long triangle-shaped interferometer (labeled as `$\Delta$-10{\rm km}'), and 2) two, 15-km long L-shaped interferometers (labeled as $2{\rm L}$-15{\rm km}), see Ref.~\cite{CoBA} for further details on the configurations.
We anticipate that the performances of a 15-km
long configuration (either triangular or 2L-shaped) are roughly a factor of two better.

\begin{figure*}
    \centering
    \includegraphics[width=0.8\textwidth]{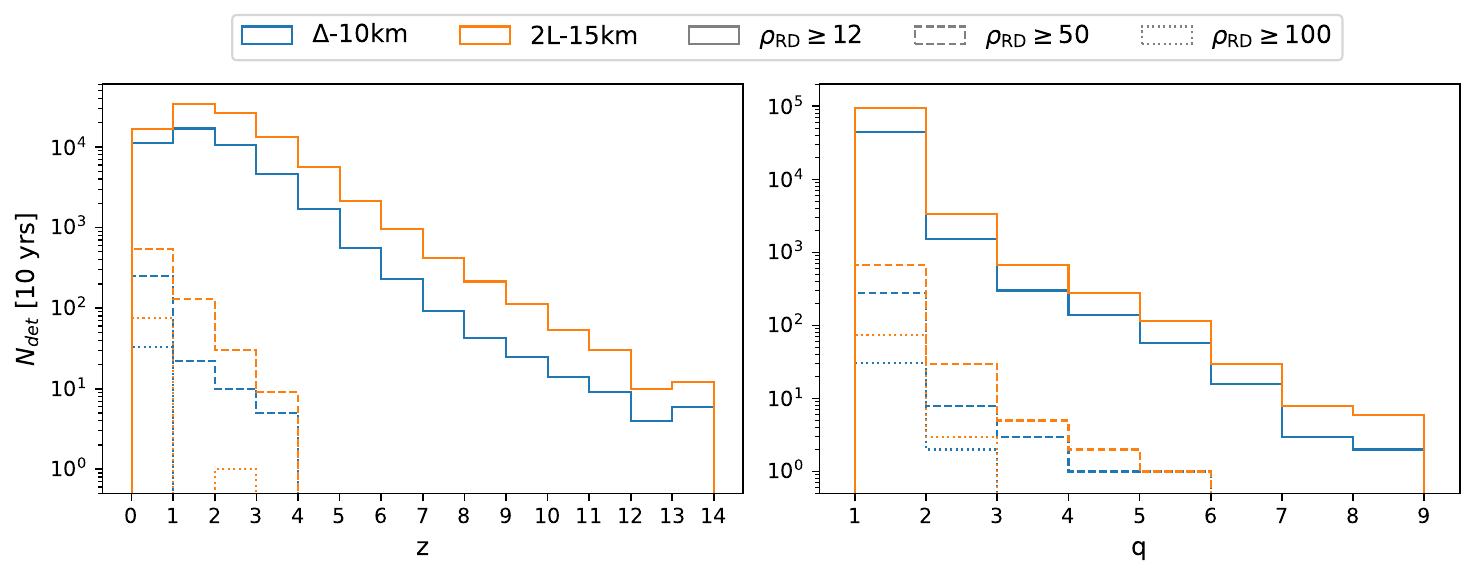}
    \caption{Left: Distribution of the number of events $N_{\rm det}$ with the redshift $z$ in the full 10-yr catalog, for small ($12$), high ($50$) and golden ($100$) ${\rho_{\rm RD}}$ thresholds. Right: Distribution of $N_{\rm det}$ with the mass ratio $q$.}
    \label{fig:ndet}
\end{figure*}

\begin{table*}[th]
    \centering
    \begin{tabular}{||c|c|c|c|c||}
         \hline\hline
         Configuration & ${\rho_{\rm RD}}\geq12$ ${\rm yr}^{-1}$& ${\rho_{\rm RD}}\geq50$ ${\rm yr}^{-1}$
         & ${\rho_{\rm RD}}\geq100$ ${\rm yr}^{-1}$ & max(${\rho_{\rm RD}}$)\\
         \hline\hline
         $\Delta$-10km & $4594\pm61$ & $28\pm7$ & $3\pm1$ & 1134\\
         2L-15km & $10071\pm88$ & $70\pm9$ & $7\pm3$ & 1262\\
         \hline\hline
    \end{tabular}
    \caption{Detection rates \textit{per year} with ${\rho_{\rm RD}}$ equal to or larger than a small (12), high ($50$) and golden ($100$) threshold, respectively, for two representative ET configurations. The last column indicates the maximum ${\rho_{\rm RD}}$ in the whole 10yr catalog.}
    \label{tab:snr-count}
\end{table*}

\begin{figure}[t]
\includegraphics[width=0.45\textwidth]{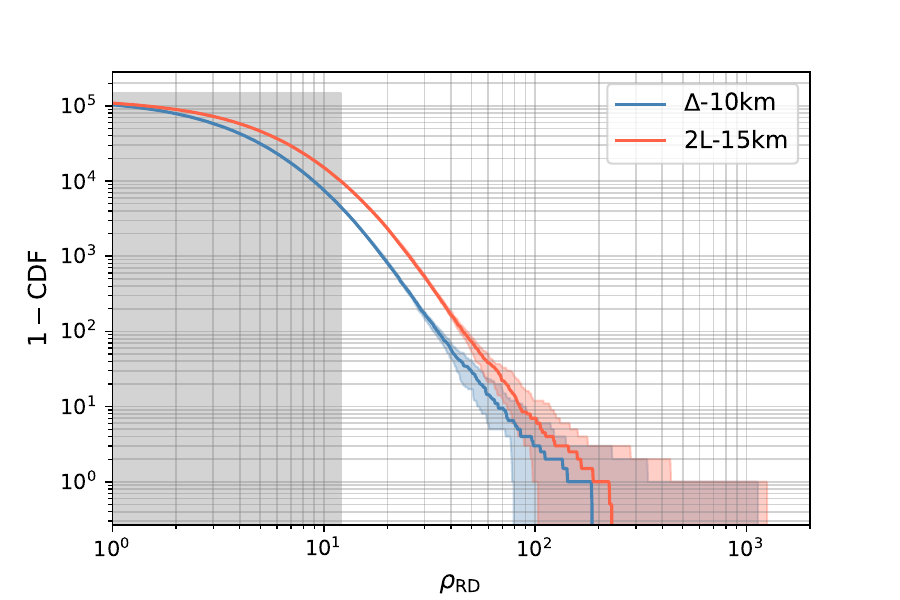}
\caption{Inverse cumulative distribution of ${\rho_{\rm RD}}$ \textit{per year} for the ET detector configurations considered in this work. The shaded gray band indicates the region where ${\rho_{\rm RD}}<12$. See also Table~\ref{tab:snr-count}.
}
\label{fig:snr-cumu}
\end{figure}

As shown in the distribution of the detected events in Fig.~\ref{fig:ndet} and from Table~\ref{tab:snr-count},  ET will detect $\sim 10^5$ BBH mergers per year, of which $\sim 4600$ (resp.~$\sim 10000$) events/yr have ${\rho_{\rm RD}}\geq 12$ in the $\Delta$-10km (resp.~2L-15km) configuration. For comparison, the GW150914 signal, which had one of the loudest ringdown signal detected by LIGO, had ${\rho_{\rm RD}}\sim 8$ when computed from $t\sim10 M$ after the peak \cite{LIGOScientific:2016lio}. Also, we anticipate the detection of a handful of events/yr with ${\rho_{\rm RD}}\geq 100$. As shown in Sec.~\ref{sec:golden}, these golden events allow us to perform precision tests with unprecedented accuracy. We summarize the median event rates and their corresponding errors across the full 10-yr catalog in Table~\ref{tab:snr-count}, and in Fig.~\ref{fig:snr-cumu} we show the inverse cumulative distribution of the ringdown SNR. As it is expected, the event rate uncertainties scale approximately as Poisson counting errors, $\sigma(N)\approx\sqrt{N}$, and depending on the particular realization of the catalog one can have individual signal-to-noise ratios as large as $\rho_{\rm RD}\sim1000$; these constitute promising candidates to identify spectra from modified gravity theories \cite{Pacilio:2023mvk}.

\begin{figure*}[t]
    \centering
    \includegraphics[width=0.45\textwidth]{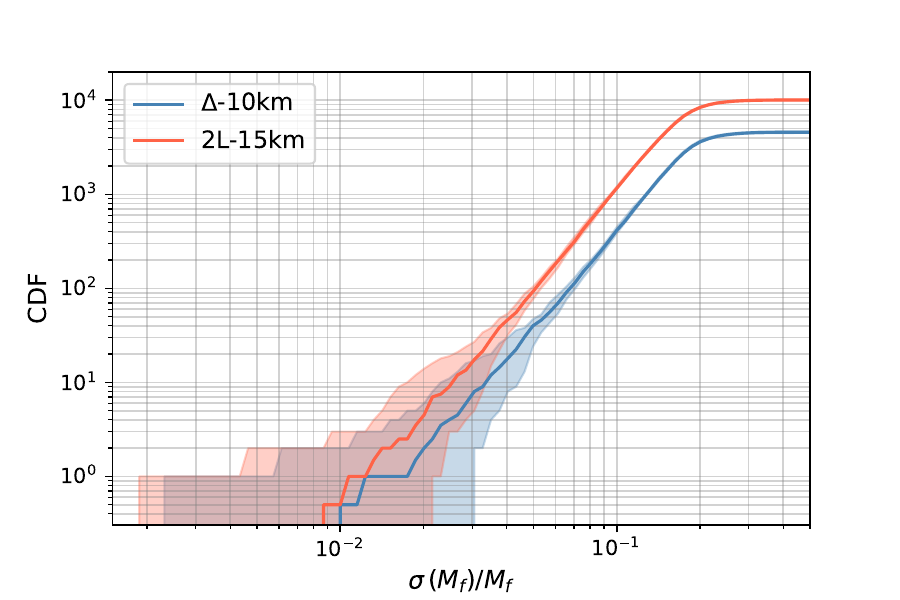}
    \includegraphics[width=0.45\textwidth]{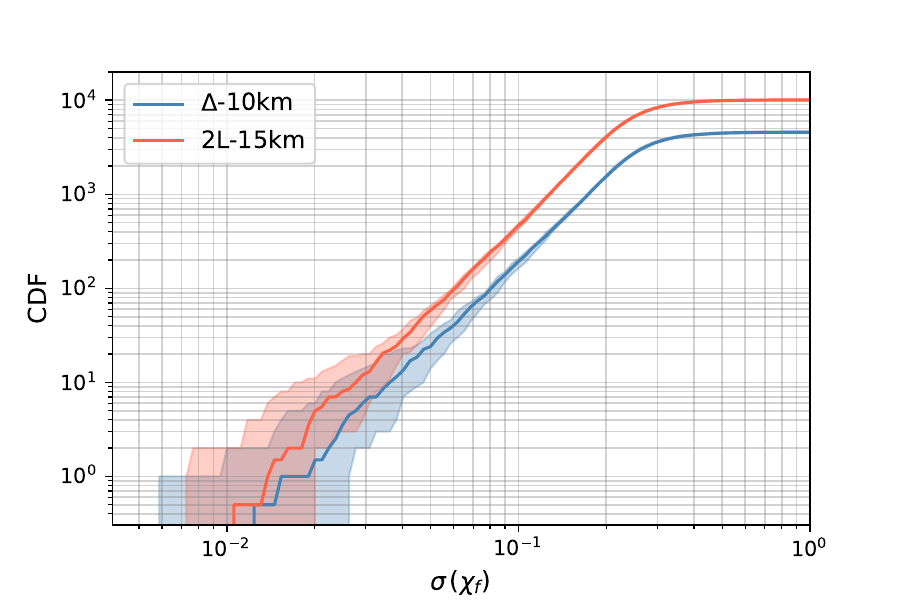}\\
    \includegraphics[width=0.45\textwidth]{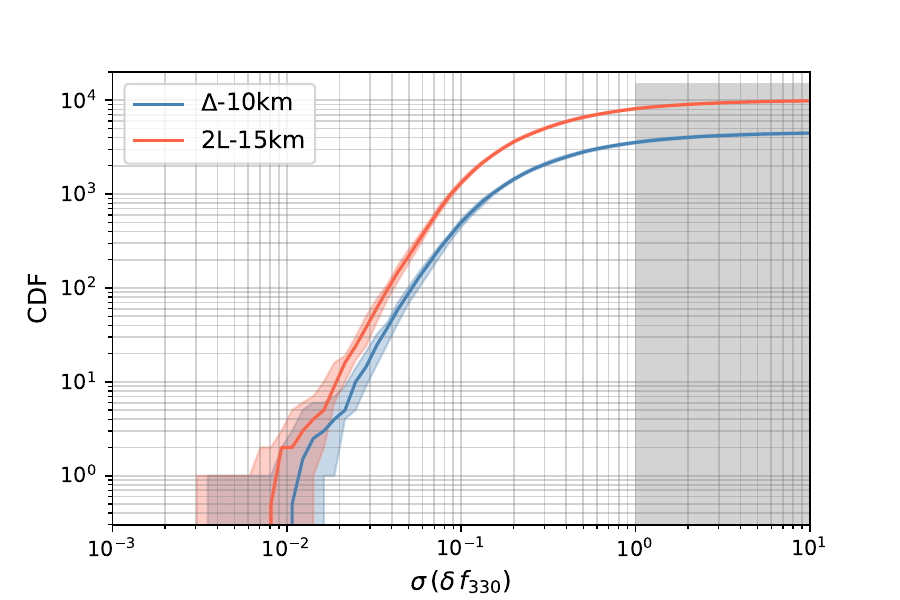}
    \includegraphics[width=0.45\textwidth]{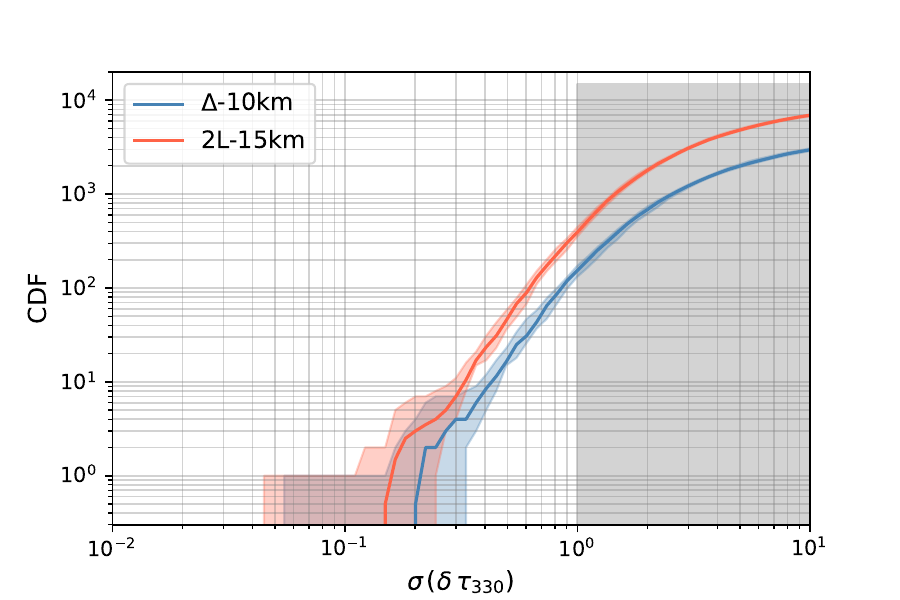}\\
    \includegraphics[width=0.45\textwidth]{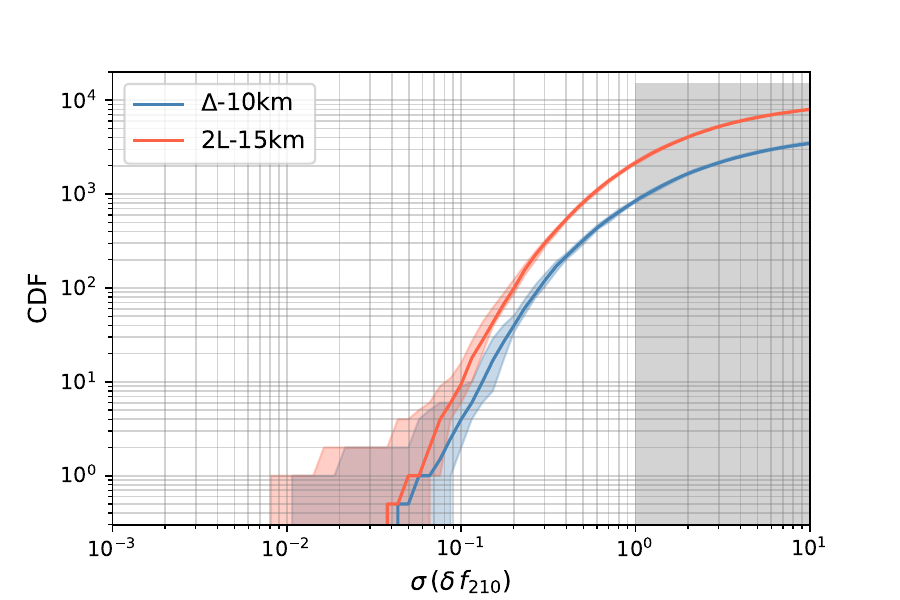}
    \includegraphics[width=0.45\textwidth]{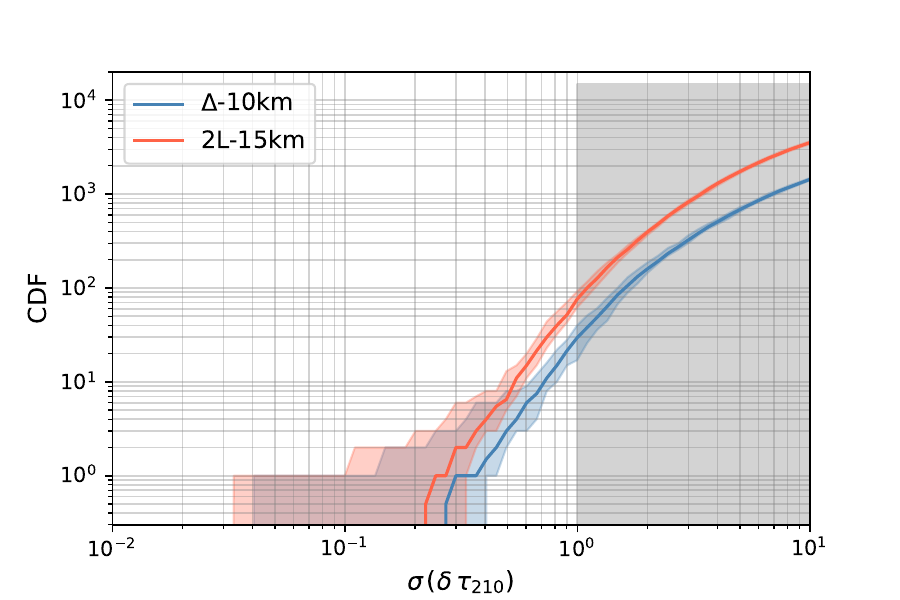} \\
    \includegraphics[width=0.45\textwidth]{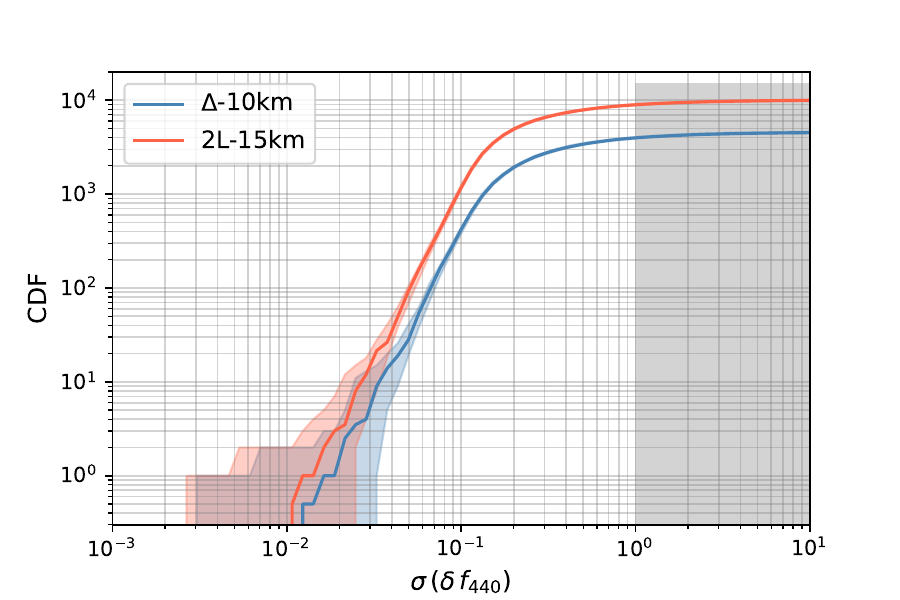}
    \includegraphics[width=0.45\textwidth]{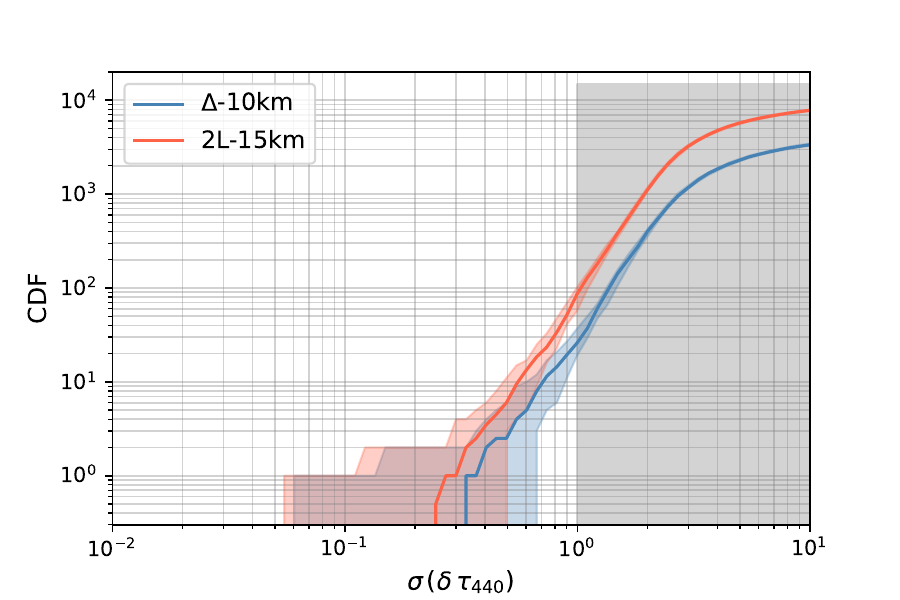}
    \caption{
    Cumulative distributions of the uncertainties in measuring various ringdown quantities (events per year for the ET configurations considered in this work).
    Top panels: $\sigma\left(M_f\right)/M_f$ (left) and $\sigma\left(\chi_f\right)$ (right).
    Second to fourth row: cumulative distributions for the error on the QNM frequency, $\sigma\left(\delta f_{lmn}\right)$ (left) and QNM damping time, $\sigma\left(\delta \tau_{lmn}\right)$ (right).
    The colored bands indicate the uncertainties on the $y$-axis computed across a 10-year catalog. The shaded vertical gray band indicates the region $\sigma(\delta f_{lmn})>1$ or $\sigma(\delta \tau_{lmn})>1$ where the constraints on $\delta f_{lmn}$ or $\delta \tau_{lmn}$ are uninformative.
    }
    \label{fig:sigma-df-cumu}
\end{figure*}

Our primary results for prospects of BH spectroscopy are summarized in Fig.~\ref{fig:sigma-df-cumu} where we show the
cumulative distributions of the uncertainties in measurements of the QNM frequencies and damping times\footnote{Here and in Fig.~\ref{fig:sigma-df-ce-cumu} we compute uncertainties only for the events that meet an SNR threshold for parameter estimation, conventionally set to $\rho_{\rm RD}\geq12$.}.
The first row corresponds to the final mass and spin of the BH; these are degenerate with $\delta f_{220}$ and $\delta \tau_{220}$. Although this does not have a direct consequence on prospects of BH spectroscopy, it gives an estimate of the performance of
consistency tests informed by the inspiral part of the signal. Note that ET will measure the final mass within fractional errors $\leq 0.1$ for $\sim 400$ (resp.~$\sim1000$) event/yr in the $\Delta$-10km (resp.~2L-15km) configuration. Similarly, for the final spin, we the absolute errors are expected to be $\leq 0.1$ for $\sim200$-$400$ event/yr.

From the second to fourth rows of Fig.~\ref{fig:sigma-df-cumu} we show the cumulative distribution for the uncertainties in measurements of the QNM frequencies (left panels) and damping times (right panels) for some of the promising subdominant modes.
Comparing the left and the right panels, we confirm that the deviation in the subdominant mode frequencies can be measured more accurately than their corresponding damping times. We expect between a few hundred and a thousand of events per year to allow for $\leq 10 \%$ uncertainty in $f_{330}$ and a handful of events per year with an uncertainty $\leq 2 \%$.  In addition, a few event/yr allow for the measurement of $f_{210}$ with an uncertainty $\leq 10 \%$, and between a few hundred and a thousand events/yr allow for the measurement of $f_{440}$ with $\leq 10 \%$ uncertainty.

Note that, while $\mathcal{A}_{440}$ is not very high, the $440$ mode can compete with the performance of the $330$ mode due to statistical abundance of the systems in which it is excited.

\section{Prospects for combined CE-ET BH spectroscopy}\label{sec:ETCE}
In this section we study the prospect of BH spectroscopy assuming a combined detection by ET and a single CE detector with 40-km arm length. As we shall see, we find that the recovery of QNM parameters is significantly improved in this case because of the increase in $\rho_{\rm RD}$.

\begin{figure*}
    \centering
    \includegraphics[width=0.8\textwidth]{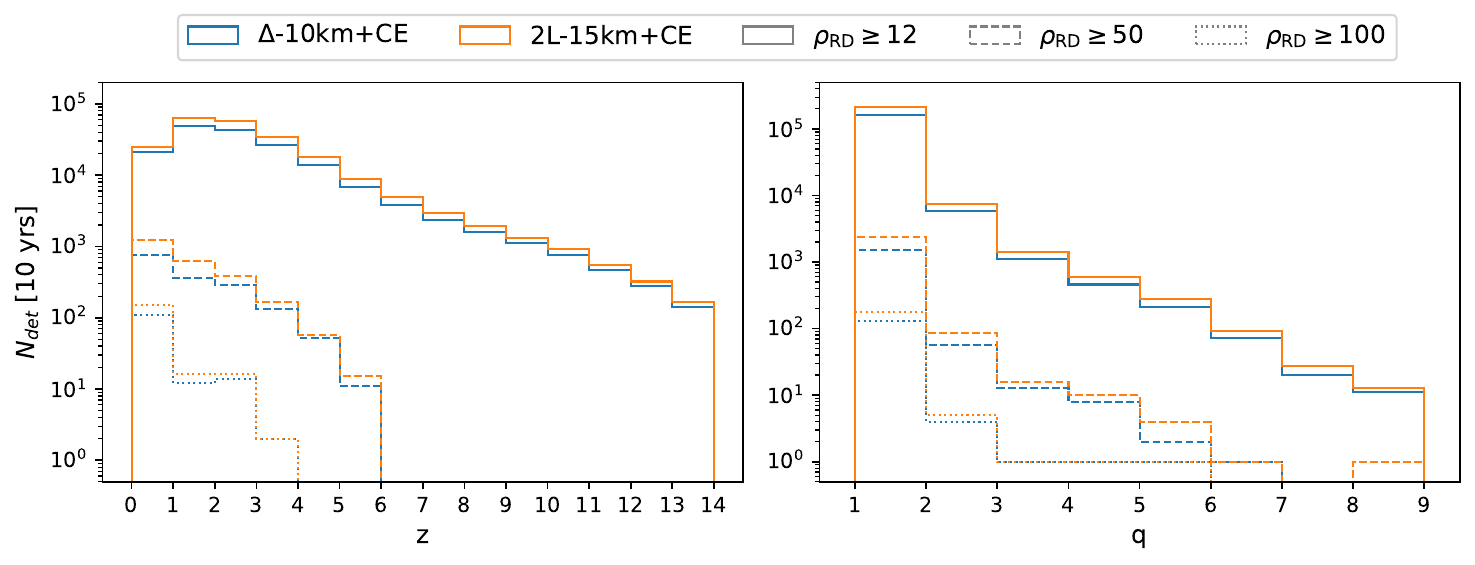}
    \caption{Same as Fig.~\ref{fig:ndet} for a network of ET operating in synergy with a single CE detector.}
    \label{fig:ndet_ce}
\end{figure*}

\begin{figure}
\includegraphics[width=0.45\textwidth]{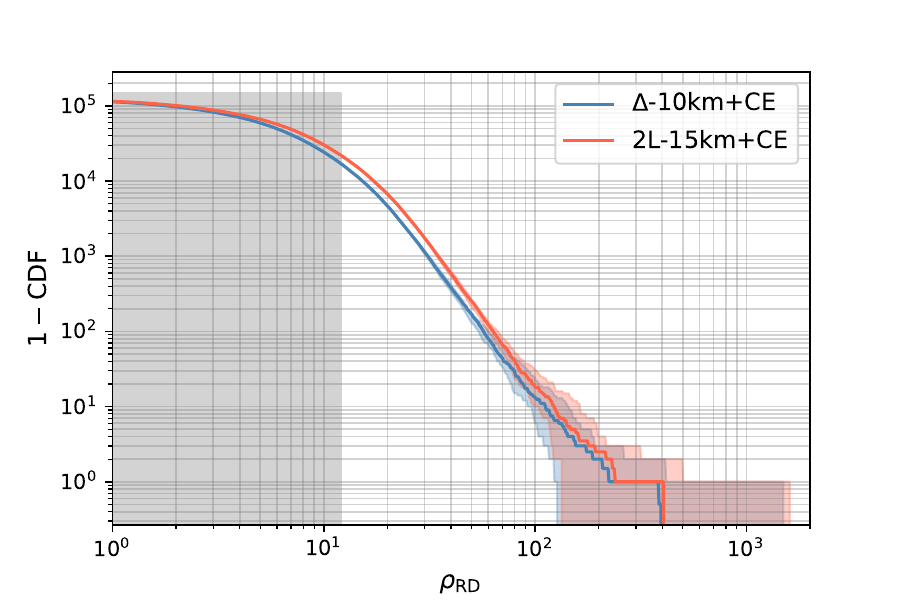}
\caption{Same as Fig.~\ref{fig:snr-cumu} but for ET operating in synergy with a single CE detector.
}
\label{fig:snr-ce-cumu}
\end{figure}

To facilitate the comparison, we will show exactly the same plots presented in Sec.~\ref{subsec:landscapeET}.
Figures~\ref{fig:ndet_ce} and \ref{fig:snr-ce-cumu} are the analog of Figs.~\ref{fig:ndet} and \ref{fig:snr-ce-cumu}, respectively, whereas Table~\ref{tab:snr-ce-count} is the analog of Table~\ref{tab:snr-count}.
One of the most striking features is that the number of events per year with ringdown SNR ${\rho_{\rm RD}}\geq100$ increases by roughly a factor 3 detected up to redshift $z\approx 6$.
Furthermore, a handful of events will have ringdown SNR of a few hundreds.

\begin{table*}[]
    \centering
    \begin{tabular}{||c|c|c|c|c||}
         \hline\hline
         Configuration & ${\rho_{\rm RD}}\geq12$ ${\rm yr}^{-1}$& ${\rho_{\rm RD}}\geq50$ ${\rm yr}^{-1}$
         & ${\rho_{\rm RD}}\geq100$ ${\rm yr}^{-1}$ & max(${\rho_{\rm RD}}$)\\
         \hline\hline
         $\Delta$-10km+CE & $17174\pm115$  & $161\pm14$ & $13\pm5$ & 1508 \\
         2L-15km+CE & $22144\pm122$ & $246\pm16$ & $18\pm7$ & 1607\\
         \hline\hline
    \end{tabular}
    \caption{Same as Table~\ref{tab:snr-count} but for ET operating in synergy with a single CE detector.}
    \label{tab:snr-ce-count}
\end{table*}

\begin{figure*}[t]
    \centering
    \includegraphics[width=0.45\textwidth]{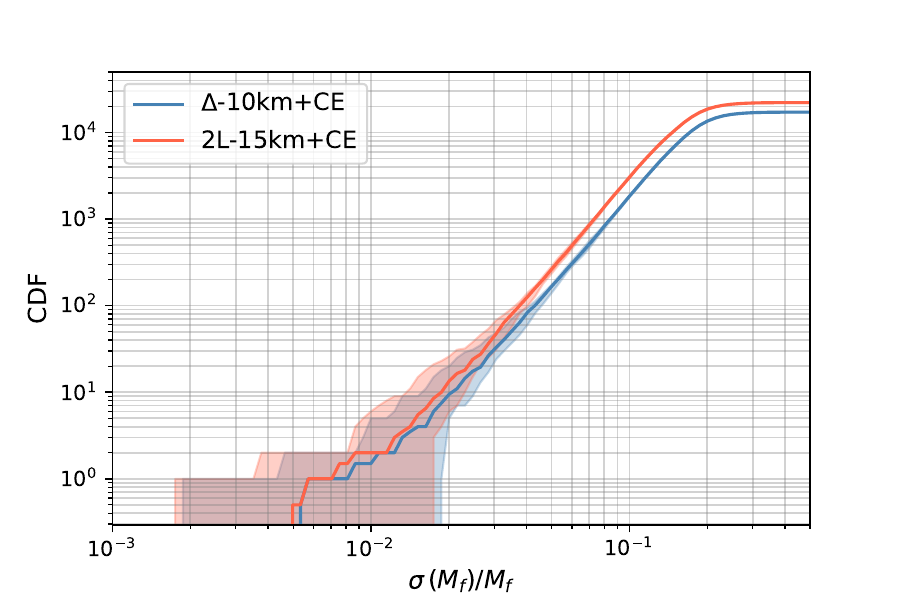}
    \includegraphics[width=0.45\textwidth]{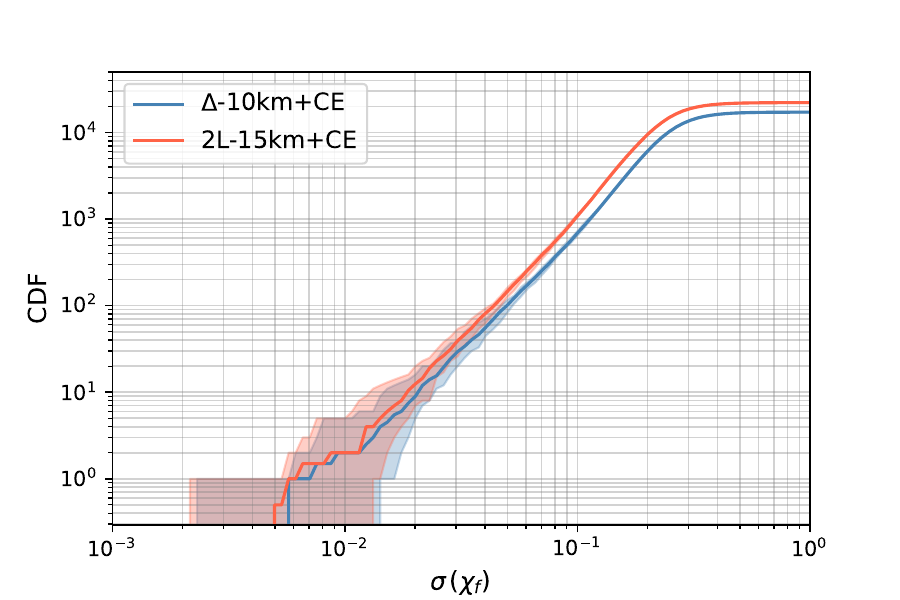} \\
    \includegraphics[width=0.45\textwidth]{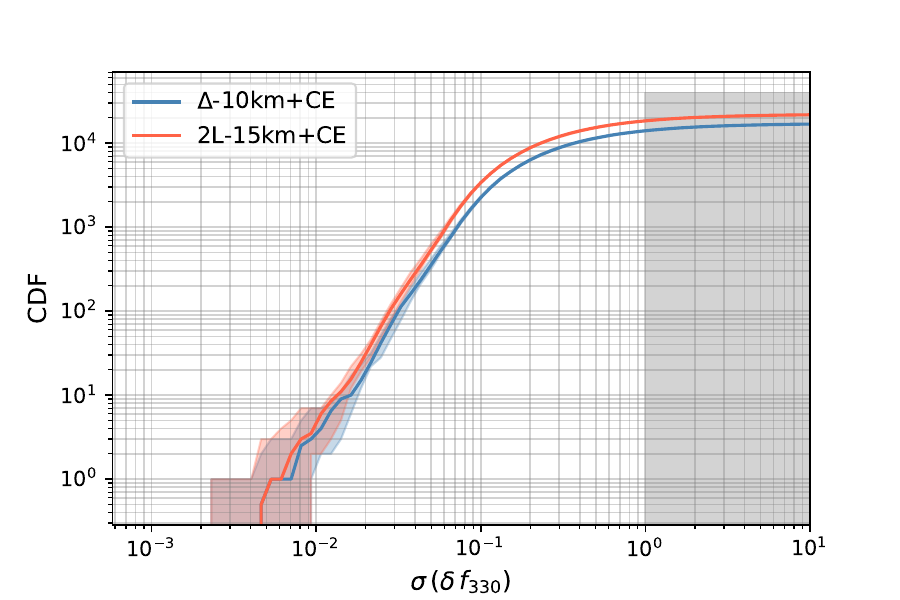}
    \includegraphics[width=0.45\textwidth]{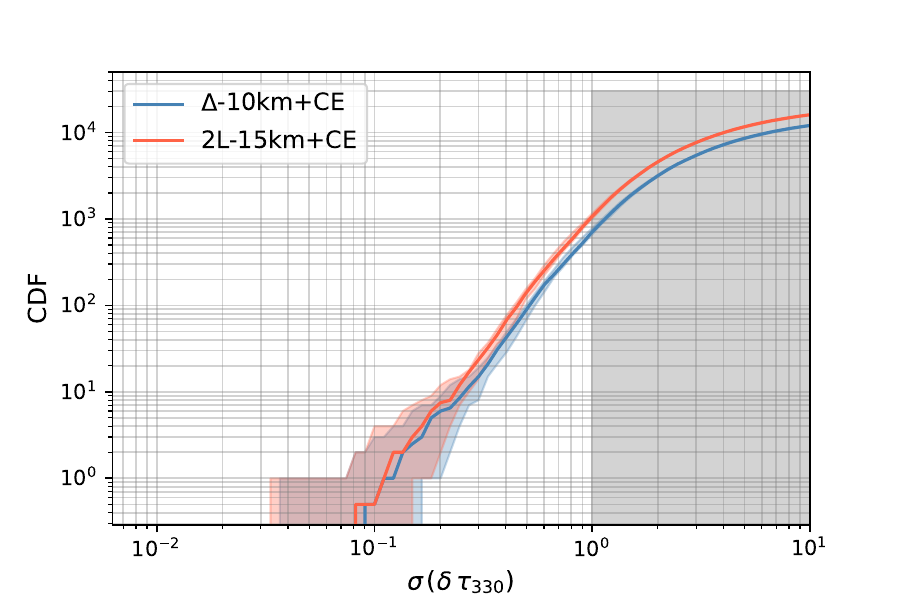} \\
    \includegraphics[width=0.45\textwidth]{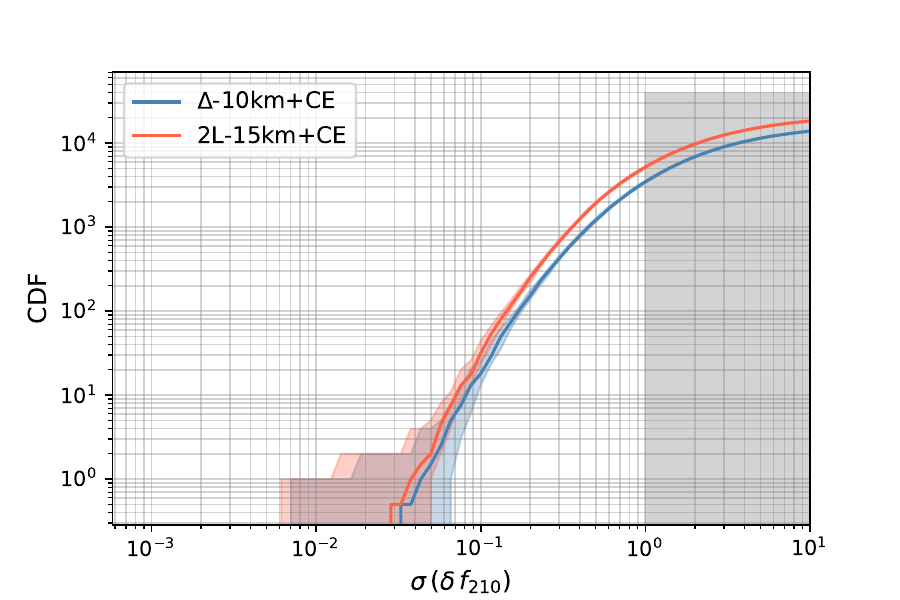}
    \includegraphics[width=0.45\textwidth]{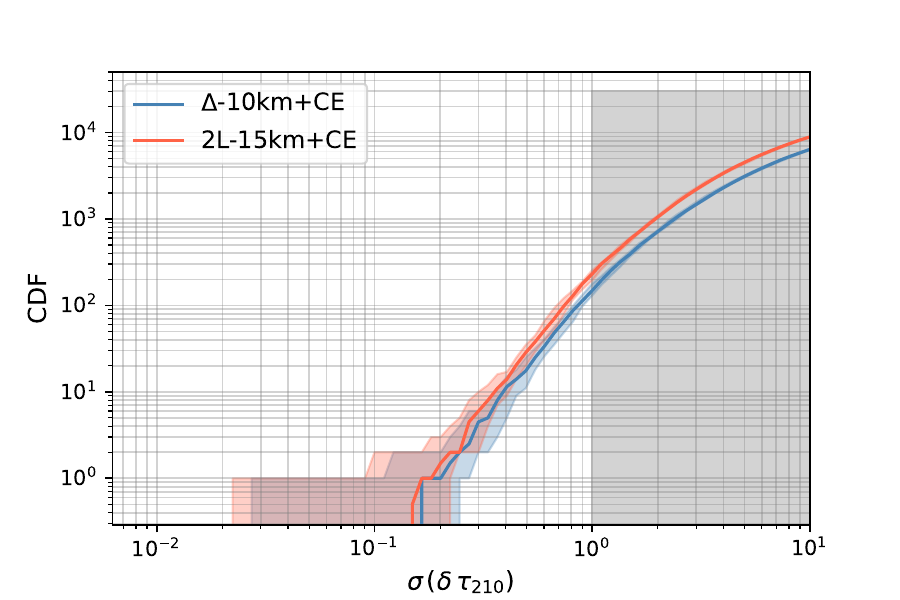}\\
    \includegraphics[width=0.45\textwidth]{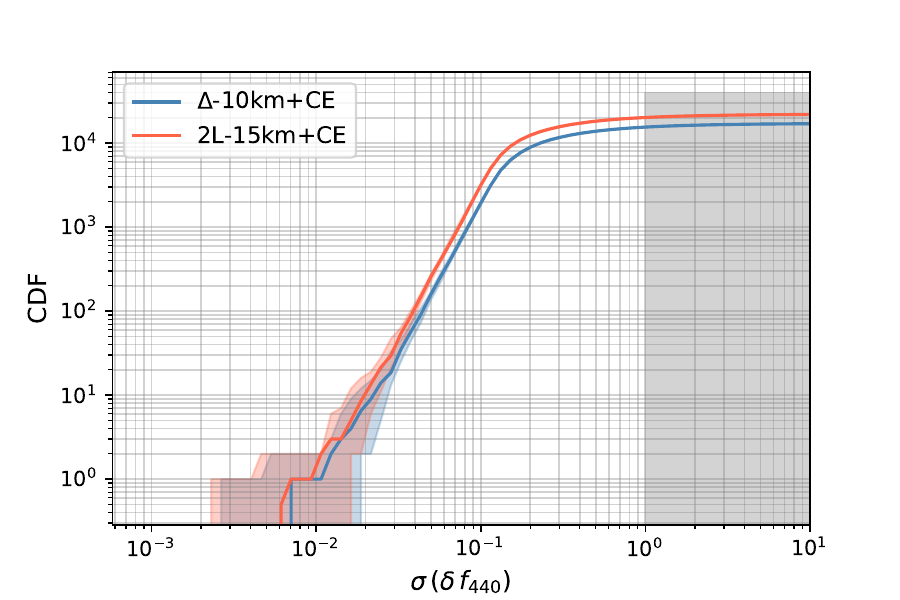}
    \includegraphics[width=0.45\textwidth]{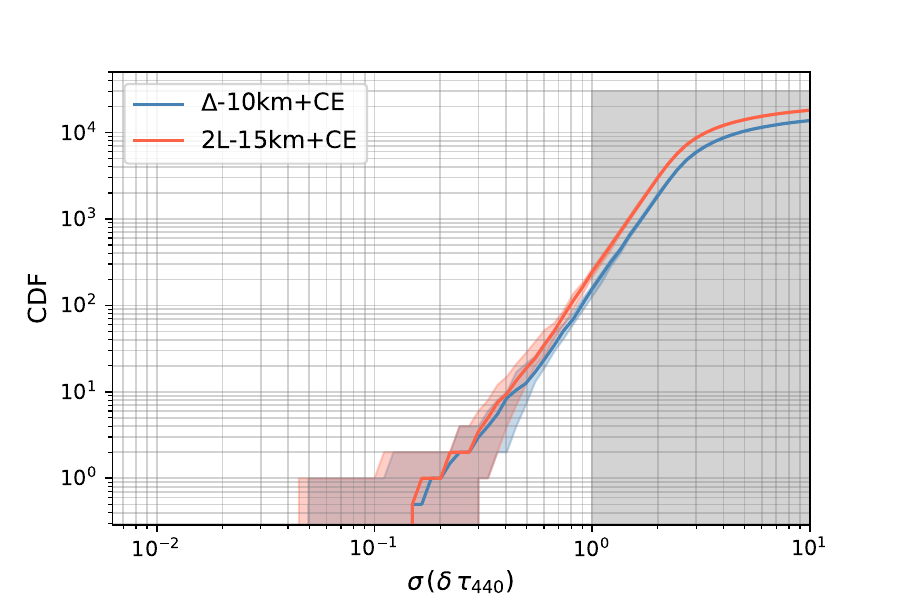}
    \caption{
    Same as Fig.~\ref{fig:sigma-df-cumu} but for ET operating in synergy with a single CE detector.
    }
    \label{fig:sigma-df-ce-cumu}
\end{figure*}

Finally, our main results for spectroscopy are again summarized in a single plot (Fig.~\ref{fig:sigma-df-ce-cumu}, which is the analog of Fig.~\ref{fig:sigma-df-cumu}).
We see that $\sim 10 $ events per year will allow  for fractional uncertainty $ \sigma (M_{f}) / M_{f}  \leq 0.02 $ (to be compared with  $\sim 1 $ event/yr for ET alone at this level of accuracy) and that few thousands event/yr will allow  for $ \sigma (M_{f}) / M_{f} \leq 10^{-1}$ (to be compared with $\sim 400$-$1000$ event/yr with ET alone).
Similar improvements are found for the measurement of the final spin.
Finally, the combined CE-ET measurements of the subdominant QNMs yield $\sim 2000$ event/yr with uncertainty $ \sigma (\delta f_{lmn}) \leq 0.1$, and $\sim 10 $ events with $\sigma (\delta f_{lmn}) \leq 0.02$ for both the 330 and the 440 modes. The accuracy in measuring $f_{210} $ performs worse and we predict $ \sigma (\delta f_{lmn}) \leq 0.3 $ for $\sim 100$ events/yr.
Looking at the right panels in Fig.~\ref{fig:sigma-df-ce-cumu}, we see that the best combined CE-ET measurement of the damping time is for $\tau_{330}$  with $\sim 10 $ event/yr allowing for $\sigma (\delta \tau_{330}) \leq \ 0.3$.

\section{Discussion and Conclusions} \label{sec:conclusions}
In this paper, we have investigated the prospects for BH spectroscopy with the ET detector in different configurations, possibly in combination with CE.
We estimated both projected bounds with isolated golden merger events and rates of accurate measurements using a state-of-the-art population model for stellar-origin BBHs informed by LIGO-Virgo-KAGRA data.
Our results highlight the importance of longer detector arms and combined CE-ET detections for what concerns ringdown tests.

Our analysis was intended to explore broadly the landscape of ringdown tests with third-generation interferometers, considering different detector configurations and networks,
and including different subdominant modes. As such, it can be extended in various ways if one wishes to perform a more detailed and focused analysis.
An obvious extension is to perform a Bayesian inference, possibly including the priors on the QNM amplitudes and phases as discussed in this work. 
Furthermore, given the large number of ringdown signals expected in the ET-CE era, a natural extension is to explore the possibility of stacking multiple signals to improve the accuracy of ringdown tests~\cite{Yang:2017zxs,Berti:2018vdi}.
For what concerns possible improvements on the ringdown modeling, it would be relevant to include the effects of mode-mixing in the amplitudes due to the expansion in spheroidal harmonics \cite{London:2018gaq} or alternatively using the parametrization in \cite{Isi:2021iql} that avoids considering mode-mixing; it would also be relevant to include overtones and quadratic effects, along the lines of the recent analysis in Ref.~\cite{Baibhav:2023clw}.
In this context, it is also relevant to note that, given the stellar-mass BBH population favored by current GW data, the 330 mode is the optimal angular mode for BH spectroscopy with ET.
This is fortunate, because the 330 mode is expected to be less contaminated by QNM mixing at the quadratic level relative to the 440 mode (which shows only slightly worse measurement accuracy). Indeed, angular-momentum sum rules imply that, within GR, the 440 mode is sourced by the dominant 220+220 mode~\cite{Cheung:2022rbm,Mitman:2022qdl,Lagos:2022otp}, whereas the 330 mode could be sourced by the 220+210 mode, whose contribution is suppressed due to the smaller excitation of the 210. Therefore, standard (linear) ringdown tests with the 330 mode should be reliable also for very loud events as those expected in the third-generation era.

\acknowledgments
We thank Emanuele Berti and Francesco Iacovelli for useful conversations and Gregorio Carullo for comments on the draft.
The research leading to these results has been conceived and developed within the ET Observational Science Board (OSB).
Numerical calculations have been made possible through a CINECA-INFN agreement, providing access to resources on MARCONI at CINECA.
S.B. would like to acknowledge the UKRI Stephen Hawking Fellowship funded by the Engineering and Physical Sciences Research Council (EPSRC) with grant reference number EP/W005727 for support during this project.
C.P.~is supported by European Union's H2020 ERC Starting Grant No. 945155--GWmining and by Cariplo Foundation Grant No. 2021-0555.
MM acknowledges financial support from the European Research Council for the ERC Consolidator grant DEMOBLACK, under contract no.~770017.
P.P. acknowledge financial support provided under the European
Union's H2020 ERC, Starting Grant agreement no.~DarkGRA--757480 and under the MIUR PRIN programme, and support from the Amaldi Research Center funded by the MIUR program ``Dipartimento di Eccellenza" (CUP:~B81I18001170001). This work was supported by the EU Horizon 2020 Research and Innovation Programme under the Marie Sklodowska-Curie Grant Agreement No. 101007855.

\newpage
\bibliography{Refs.bib}
\end{document}